\documentstyle[12pt,fleqn]{article}

\def\Slash#1{{\rm\ooalign{\hfil$/$\hfil\crcr \hbox{$#1$}}}}
\newcommand\Kmbf[1]{\mbox{\boldmath{$#1$}}}

\begin{document}

\begin{flushright}
 TIT/HEP-391/NP
\end{flushright}

\begin{center}
{\Large \bf
Dynamical Chiral Symmetry Breaking in Effective \par
Models of QCD in the Bethe--Salpeter Approach} \par
\vskip 10mm
K. Naito\footnote{E-mail address: kenichi@th.phys.titech.ac.jp},
K. Yoshida, 
Y. Nemoto, 
M. Oka\\ 
{\it Department of Physics, Tokyo Institute of Technology,} \\
{\it Meguro, Tokyo 152-8551, Japan} 
\vskip 5mm
M. Takizawa\\ 
{\it Laboratory of Computer Sciences, Showa College of Pharmaceutical Sciences,} \\
{\it Machida, Tokyo 194-8543, Japan}
\end{center}
\vskip 5mm
\begin{abstract}
\baselineskip 1.5pc
 Dynamical breaking of chiral symmetry
in effective models of QCD is studied.
Introducing a cut-off function or a non-local interaction,
the Noether current must be modified and thus 
the Ward--Takahashi identity and the PCAC relation
are modified accordingly.
We point out that the pion decay constant must be defined consistently
with the Noether current so that the low-energy relations are satisfied.
We define the proxy of the Noether current for general effective models,
which is consistent with loop expansion of the Cornwall--Jackiw--Tomboulis 
effective action. A general formula for the pion decay constant in terms
of the Bethe--Salpeter amplitude is derived.
 The effective Pagels--Stokar formula is proposed which
is useful to estimate the decay constant without solving the 
Bethe--Salpeter equation.
\end{abstract}

\newpage
\renewcommand{\thefootnote}{ \mbox{*}\arabic{footnote} }


\section{Introduction} \label{SEC:H100220:1}
 The program to derive the observed properties of hadrons
non-perturbatively
in quantum chromodynamics(QCD) has been pursued with great intensity but 
not accomplished yet. The concept of chiral symmetry and its spontaneous 
breakdown are among the most important aspects of low-energy hadron physics.
The spontaneous breakdown of chiral symmetry is believed to be responsible
for a large part of the low-lying hadron masses as well as for the emergence
of octet pseudo-scalar mesons as Nambu--Goldstone(NG) bosons. 
In order to explain
the observed hadron spectrum, one also needs small, explicitly chiral
symmetry breaking terms, namely, the flavor dependent current quark
mass terms. 

 Dynamical chiral symmetry breaking(DCSB) of QCD
has been extensively studied in effective models in terms of light
quarks.
The Nambu--Jona-Lasinio(NJL) model\cite{NJL}, as an example, is a
superb model,
that presents the main idea of DCSB most concisely.
It describes the spectrum and properties of NG mesons
fairly well.\cite{KH1}
The NJL model, however, has several shortcomings.
It may be too simple to represent some of the key properties of QCD.
First, it does not describe color confiniement and therefore
the meson spectrum above $2M_Q$ threshold ($M_Q$ is the constituent 
quark mass) is not reproduced.
 Secondly, the quark interactions are not asymptotically free in NJL.
This property will be crucial to describe short distance structure
of mesons.

 There have been various attempts to derive effective
quark theories which are consistent with the asymptotic behavior and
low-energy chiral
symmetry breaking. One of such models is the improved ladder
approximation proposed by Higashijima and Miransky.\cite{H84,M84}
This model is consistent with the asymptotic
behavior of the leading order renormalization
group analysis of QCD.\cite{Pol76}
It also realizes DCSB, as is seen in the non-zero quark mass function
given as the solution of the Schwinger--Dyson(SD) equation.
Then according to the NG theorem, the Bethe--Salpeter(BS)
equation gives a massless NG boson
 with $J^{PC}=0^{-+}$.\cite{KG1}
 The numerical results for the
pion decay constant $f_\pi$ and the quark condensate
$\langle \overline{\psi}\psi \rangle$ are consistent with
 other analyses.\cite{GL82}
On the other hand,
 it is known that this approximation violates the axial-vector
Ward--Takahashi identity.\cite{KG2}
Therefore it looks unsuitable to study the low-energy relations in
QCD.

 However, there exists a global chiral symmetry in the improved
ladder approximation. It produces 
a Noether current associated with the axial transformation, 
namely axial-vector current $J^\alpha_{5\mu}$, 
which is generally modified from the QCD
form $\displaystyle{\overline{\psi}
\gamma_\mu\gamma_5\frac{\lambda^\alpha}{2}\psi}$. 
 Accordingly the pion decay constant defined by
\begin{equation}
 ip_\mu f_\pi \delta_{\alpha\beta} := \langle 0 | 
 J^\alpha_{5\mu}(0)|\pi^\beta(p)\rangle
 \label{AEQ:H100311:1}
\end{equation}
must be modified. The Ward--Takahashi identity for the modified 
axial-vector current is satisfied. 

The improved ladder approximation is considered as
a typical example of the effective model
which contains a non-local interaction term, while 
various similar models are 
proposed.\cite{KA,KG3,NUCL1994,BQRT,MR,KG4,nonlocalNJL,Instanton,Dgl,HRW,BA1}
Among these effective models
some require no modification of the 
axial-vector current. However, 
the corresponding Bethe--Salpeter equations become so complicated
that practical calculation 
is rather difficult in such models.
In Ref.\cite{KA}, the truncation in the 
Gegenbauer polynomial expansion is employed. 
This procedure may violate the 
low-energy relations. In Ref.\cite{KG3}, to preserve the axial-vector
Ward--Takahashi identity, the running gauge parameter is employed,
which makes practical calculation hard for finite quark mass.
In Ref.\cite{MR}, the SD and BS equations are solved consistently. 
But as the wave function renormalization constant $Z_2$ deviates from 1,
the vector current or the charge operator should be modified.

 On the other hand, the improved ladder
approximation is simple and useful to treat numerically.
 The aim of this paper is to study a systematic approach which 
repairs the low-energy relations in
the effective models and makes it possible to study the low-energy
relations in QCD.
 We propose to define the axial-vector current consistenly
with the approximation in the effective model, and describe the low-energy
relation in terms of the solutions of the SD and BS equations. We derive a
formula for the consistent pion decay constant.
There have been similar studies of this problem
in the literature\cite{KG4,nonlocalNJL}
where the NJL models with smooth cut-off regularization
are studied.
However our approach is more general and systematic.
We study the SD and BS equations in
 the loop expansion using the Cornwall--%
Jackiw--Tomboulis(CJT) effective action formulation.\cite{CJT} We show that
the BS equation has a massless bound state solution corresponding 
to the NG boson.
We also give a formula for the decay constant which is consistent 
with the rainbow--ladder approximation.\par
 In Sec.\ref{SEC:H100326:1}, we present how the Noether current is modified
due to the loop momentum cut-off and/or the non-local interaction. The 
modification of the axial Ward--Takahashi identity is also discussed.\par
 In Sec.\ref{SEC:H100301:1}, we employ the CJT
effective action to study the consistency of the SD and the BS equations.
The NG theorem and the pion decay constant are studied in terms of the
effective action. We further consider the explicit chiral symmetry
breaking due to the finite quark mass. Three cases in the patterns of
the local chiral symmetry breaking of the effective models are
studied in detail.\par
 In Sec.\ref{SEC:H100409:1}, we present a general formula of the pion 
decay constant in the loop expansion of the CJT acion. The formula is
given in terms of the quark full propagator as the solution of the
SD equation and the pion BS amplitude.
 We also propose a Pagels--Stokar type formula, which gives the pion
decay constant in terms only of the quark full propagator,
namely, the mass function $B(q^2)$.\par
 In Sec.\ref{SEC:H100326:3}, we employ a simple numerical model
 to check our analytical results. We confirm that the modifications of  
axial-vector current and the pion decay constant are significantly large,
while our new Pagels--Stokar type formula gives a good approximation.\par
  A conclusion is given in Sec.\ref{SEC:H100327:1}.


\section{Noether current and Low-energy relations}
\label{SEC:H100326:1}

 Before the disscussion about the chiral symmetry and low-energy
relations in the approximated SD and BS equations, we consider 
general aspects of chiral symmetry and low-energy relations
in effective models. In this section, we show how to modify 
the Noether current and the Ward--Takahashi identity from the 
original form in QCD.
This gives a perspective in the following discussion for the case of
the rainbow--ladder approximated SD and BS equations.

 We consider a general effective model of QCD whose 
lagrangian density is given by
\begin{eqnarray}
 {\cal L}[\psi,\overline{\psi}] & := & 
{\cal L}_{\rm free}[\psi,\overline{\psi}] +
 {\cal L}_{\rm int}[\psi,\overline{\psi}] \label{AEQ:H100308:1} \\
 {\cal L}_{\rm free}[\psi,\overline{\psi}] & := & \overline{\psi} 
f(\partial^2) (i\Slash{\partial}-m_0) \psi \label{AEQ:H100308:2}
\end{eqnarray}
where the quark field $\psi$ is a column vector in the color, the flavor and 
the Dirac space.
Here the function $f(\zeta)$ of $\zeta=\partial^2$ is introduced as a cut-off
regularization function in order to regularize the ultraviolet 
divergences coming from quark loops.
The reason we introduce the cut-off function at the lagrangian level
is to preserve the consistency of the SD and BS equations. If one
uses the regularization that is inconsistent between the SD and BS 
equations, the low-energy relations based on the chiral symmetry should
be violated by the regularization.
The function $f(\zeta)$ should satisfy $f(\zeta=0)=1$. 
For $\zeta \gg \Lambda_{\rm UV}^2$,
$f(\zeta)$ should diverge sufficiently fast so as to regularize loop integrals.
For example, the sharp cut-off is given by
\begin{equation}
 f(\zeta) = 1 + M \theta(\Lambda_{\rm UV}^2-\zeta),\quad M\to \infty
 \label{AEQ:H100812:1}
\end{equation}
 and the Gaussian smooth cut-off is given by
\begin{equation}
 f(\zeta) = 1 + M \exp\left( \frac{\zeta}{\Lambda_{\rm UV}^2} \right).
 \label{AEQ:H100812:2}
\end{equation}
A caveat of this procedure is that a general choice of $f(\zeta)$ may cause
a difficulty in canonical quantization. 
Since our purpose is not to study the non-local field theory,
we simply employ the path integral formulation with the action integral
regularized by $f(\zeta)$.
Although this treatment is not rigorous, it is sufficent in the present
disscussion.    
 We thus maintain the consistency between the
chiral symmetry and the regularization.\par
 $f(\zeta)$ also determines a scale $\Lambda_{\rm UV}$ 
at which the bare quark mass $m_0$
is evaluated.\cite{KG1,NUCL1994}\,
To compare the bare mass $m_0$ the renormalized quark mass $m_R$ 
in QCD, 
one must impose a suitable renormalization condition.
In general $m_0$ is a diagonal flavor matrix, i.e., 
$m_0={\rm diag}(m_u,m_d,m_s)$
for $N_f=3$. But in this paper we deal only with
a flavor independent mass or the $SU(3)$ limit.
The generalization to the flavor dependent masses can be also done.
Although there is a difficulty in
choosing the center of mass coordinate for the quark--antiquark 
bound states physically, the low-energy relations hold for any choice
of the center of mass coordinate.\cite{MR} \par
A general non-local 4-quark interaction is written by
\begin{eqnarray}
  {\cal L}_{\rm int}[\psi,\overline{\psi}](x) &:=& -\frac{1}{2} \int_{pp'qq'} {\cal K}^{mm',nn'}(p,p';q,q') \nonumber \\
 & & {} \times \overline{\psi}_m(p)\psi_{m'}(p') \overline{\psi}_n(q) \psi_{n'}(q') e^{-i(p+p'+q+q')x}
 \label{AEQ:H100308:3}
\end{eqnarray}
where $\int_p$ denotes $\int \frac{d^4 p}{(2\pi)^4}$ and
the Fourier transformation of the quark field is defined by
\begin{eqnarray}
\overline{\psi}(p) & := & \int d^4x e^{ipx}\overline{\psi}(x),
        \label{AEQ:H100318:1} \\
 \psi(p) & := & \int d^4x e^{ipx}\psi(x).
        \label{AEQ:H100318:2}
\end{eqnarray}
The indices $m,n,\cdots$
are combined indices $m:=(a,i,f),\,n:=(b,j,g),\cdots$ with Dirac indices
$a,b,\cdots$ and color indices $i,j,\cdots $ and flavor indices $f,g,\cdots$.
We do not consider the quantum correction and renormalization 
for the interaction itself, 
because we assume that effective models have already contained such effects. 
${\cal K}^{mm',nn'}(p,p';q,q')$ is an interaction kernel which we can 
assume without loss of generality 
\begin{equation}
 {\cal K}^{mm',nn'}(p,p';q,q') = {\cal K}^{nn',mm'}(q,q';p,p').
 \label{AEQ:H100308:4}
\end{equation}
We further assume that the interaction is 
invariant under the global $SU(3)_L\times SU(3)_R$ transformation
\begin{eqnarray}
 {\cal K}^{mm',nl}(p,p';q,q') (i\frac{\lambda^\alpha}{2})_{ln'} & = &
 (i\frac{\lambda^\alpha}{2})_{nl} {\cal K}^{mm',ln'}(p,p';q,q') 
 \label{AEQ:H100318:3} \\
 {\cal K}^{mm',nl}(p,p';q,q') (i\gamma_5\frac{\lambda^\alpha}{2})_{ln'} & = &
 -(i\gamma_5\frac{\lambda^\alpha}{2})_{nl} {\cal K}^{mm',ln'}(p,p';q,q') 
 \label{AEQ:H100308:5}
\end{eqnarray}
 where $\lambda^\alpha$ is the Gell-Mann matrix
$(\alpha=1,\cdots,8)$ for $N_F=3$
with normalization condition
${\rm tr}^{\rm (F)}[\lambda^\alpha \lambda^\beta]=2 \delta^{\alpha\beta}$.
Thus the violation of the global chiral symmetry comes 
from the quark mass term only. 
We here concentrate only on the chiral symmetric 
4-quark interaction for simplicity.
The generalization to the multi-quark interaction is
straightforward.\cite{TIT02}
Some effective models in Ref.\cite{U1A,Instanton,TIT02} contain the 
interaction which violates the global $U(1)_A$ symmetry, and our
approach can be generalized to such models as well.

 Under the infinitesimal axial transformation
\begin{equation}
 \psi(x) \to \psi'(x) := (1+i\gamma_5 \frac{\lambda^\alpha}{2} \theta^\alpha(x)) \psi(x), \label{AEQ:H091004:2} 
\end{equation}
the action
\begin{equation}
 S[\psi,\overline{\psi}]:= \int d^4 x {\cal L}[\psi,\overline{\psi}](x)
\label{AEQ:H100308:6}
\end{equation}
changes by
\begin{eqnarray}
\lefteqn{ \Delta_5 S[\psi,\overline{\psi}] :=S[\psi',\overline{\psi}{}'] - S[\psi,\overline{\psi}] } \nonumber \\
 & \equiv  &
  \int d^4x \theta^\alpha(x) \Bigg\{ \overline{\psi}(x)(\stackrel{\leftarrow}{
  \Slash{\partial}} f(\stackrel{\leftarrow}{\partial}{}\!^2)\gamma_5 \frac{
  \lambda^\alpha}{2} + f(\partial^2)\Slash{\partial}\gamma_5\frac{
  \lambda^\alpha}{2}) \psi(x)  \nonumber \\
& & {} \quad - 2m_0 i \overline{\psi}(x) \frac{f(\stackrel{\leftarrow}{
\partial}{}\!^2)+f(\partial^2)}{2}  \gamma_5 \frac{\lambda^\alpha}{2} \psi(x) 
 \label{AEQ:H100308:7} \\
 & & -\int_{pp'qq'} \bigg( {\cal K}^{mm',nn'}(p,p';-p-p'-q',q') - {\cal K}^{mm',nn'}(p,p';q,-p-p'-q) \bigg) \nonumber \\
 & & \times \overline{\psi}_m(p) \psi_{m'}(p') \overline{\psi}_n(q) (i\gamma_5\frac{\lambda^\alpha}{2}\psi)_{n'}(q') e^{-i(p+p'+q+q')x} \Bigg\}. \nonumber
\end{eqnarray}
 One finds that the third term in the right hand side(RHS) of
Eq.$(\ref{AEQ:H100308:7})$
vanishes if the momentum dependence of the interaction kernel
 is such that   
\begin{equation}
 {\cal K}^{mm',nn'}(p,p';q,q') = {\cal K}^{mm',nn'}(p+p';q+q').
 \label{AEQ:H100308:8}
\end{equation}
This relation is satisfied when the kernel is generated by one gluon
exchange whose coupling depends only on the transfer
momentum.\cite{KA,KG3,BQRT,MR}
 Furthermore if the cut-off function $f(\zeta)$ is identically $1$, 
we obtain the operator identity
\begin{equation}
 \partial^\mu J^\alpha_{5\mu}(x) = 2m_0 J^\alpha_5(x)
 \label{AEQ:H100308:9} 
\end{equation}
 for
\begin{eqnarray}
 J^\alpha_{5\mu}(x) & := & \overline{\psi}\gamma_\mu\gamma_5
 \frac{\lambda^\alpha}{2}\psi(x), \label{AEQ:H100308:9b} \\
 J^\alpha_5(x) & := & \overline{\psi}i\gamma_5\frac{\lambda^\alpha}{2}\psi(x).
 \label{AEQ:H100308:10}
\end{eqnarray}
For general ${\cal K}^{mm',nn'}(p,p';q,q')$ and $f(\zeta)$,
Eq.$(\ref{AEQ:H100308:7})$ is written as 
\begin{equation}
  \Delta_5 S[\psi,\overline{\psi}] \equiv \int d^4x \theta^\alpha(x) \bigg\{
  \partial^\mu \tilde{J}^\alpha_{5\mu} - 2m_0 I^\alpha_5 \bigg\}(x)
  \label{AEQ:H100308:11}
\end{equation}
with
\begin{eqnarray}
 \tilde{J}_{5\mu}^\alpha(x) & := & I_{5\mu}^\alpha(x) - K_\mu^\alpha(x),
 \label{AEQ:H100308:11a} \\
 I_{5\mu}^\alpha(x) & := & \int_{p,q} \overline{\psi}(p) \Big( \gamma_\mu
 f_1(-p^2,-q^2) \nonumber \\
 & & {} \quad + (p-q)_\mu (\Slash{p}-\Slash{q}) f_2(-p^2,-q^2) \Big)
 \gamma_5 \frac{\lambda^\alpha}{2} \psi(q) e^{-i(p+q)x},
 \label{AEQ:H100308:12} \\
 K_\mu^\alpha(x) & := & \int_{pp'qq'} \frac{i(p+p'+q+q')_\mu}{(p+p'+q+q')^2}
 \nonumber \\
 & & \times \bigg( {\cal K}^{mm',nn'}(p,p';-p-p'-q',q')
        -{\cal K}^{mm',nn'}(p,p';q,-p-p'-q)\bigg) \nonumber \\
 & & \times \overline{\psi}_m(p)\psi_{m'}(p')
  \overline{\psi}_n(q)(i\gamma_5\frac{\lambda^\alpha}{2}\psi)_{n'}(q')
  e^{-i(p+p'+q+q')x}, \label{AEQ:H100308:14} \\
  I_5^\alpha(x) & := & \overline{\psi}(x) 
  \frac{ f(\stackrel{\leftarrow}{\partial}{}\!^2) + f(\partial^2) }{2} 
  i\gamma_5\frac{\lambda^\alpha}{2}\psi (x),
 \label{AEQ:H100308:13} 
\end{eqnarray}
\begin{equation}
 f_1(-p^2,-q^2) := \frac{f(-p^2)+f(-q^2)}{2}, \quad 
 f_2(-p^2,-q^2) := \frac{f(-p^2)-f(-q^2)}{2(p^2-q^2)}.
  \label{AEQ:H091007:2}
\end{equation}
 $\tilde{J}^\alpha_{5\mu}(x)$ is the Noether current
associated with the axial transformation Eq.$(\ref{AEQ:H091004:2})$
and is conserved in the chiral limit $m_0\to 0$. 
 For non-zero $m_0$, it satisfies the operator identity
\begin{equation}
 \partial^\mu \tilde{J}_{5\mu}^\alpha(x) = 2m_0 I^\alpha_{5}(x).
 \label{AEQ:H100318:4}
\end{equation}
Because the effective lagrangian contains non-standard momentum
dependencies, the Noether axial-vector current is non-local and
cannot be calculated in the standard procedure. To avoid confusion,
we call Eq.$(\ref{AEQ:H100308:9b})$ the {\em naive} axial-vector
current and call Eq.$(\ref{AEQ:H100308:11a})$ the {\em true}
axial-vector current.

 When chiral symmetry is broken dynamically, the Schwinger--Dyson
equation has a solution with non-zero quark mass function and
the Bethe--Salpeter equation gives a massless pion state which
must appear as a pole of the Noether current according to the NG theorem. 
We define the {\em effective}
decay constant $\tilde{f}_\pi$ associated with the Noether current by
\begin{equation}
 iP_{B\mu} \tilde{f}_\pi := \langle 0 | 
 \tilde{J}^\alpha_{5\mu}(0)| \Kmbf{P} \rangle
\label{AEQ:H100225:1}
\end{equation}
which is compared to the {\em naive} decay constant $f_\pi$ defined by
\begin{equation}
 iP_{B\mu} f_\pi := \langle 0 | 
 J^\alpha_{5\mu}(0)|  \Kmbf{P} \rangle
\label{AEQ:H100301:2}
\end{equation}
where $|\Kmbf{P}\rangle$ denotes a pion state with normalization 
condition $\langle \Kmbf{P} | \Kmbf{P}' \rangle = (2\pi)^3 2P_{B0}
\delta^3(\Kmbf{P}-\Kmbf{P}')$ and $P_{B\mu}:=(\sqrt{M_\pi^2+\Kmbf{P}^2},
\Kmbf{P})$ denotes the on-shell momentum.
The matrix element of Eq.$(\ref{AEQ:H100318:4})$ between a pion state
$\langle \Kmbf{P} \, |$ and the vacuum $|0 \rangle$ gives
\begin{equation}
 M_\pi^2 \tilde{f}_\pi = -2m_0 i 
 \int_q \frac{f(q_{+B}^2)+f(q_{-B}^2)}{2}
 {\rm tr}[\overline{\chi}(q;P_B) \gamma_5 \frac{\lambda^\alpha}{2}]
 \label{AEQ:H100301:1}
\end{equation}
where
\begin{equation}
 q_{-B} := q-\frac{P_B}{2},\, q_{+B} := q+\frac{P_B}{2}
 \label{AEQ:H100318:5}
\end{equation}
and the BS amplitude $\chi(q;P_B)$ and its conjugate are
defined by
\begin{eqnarray}
 \chi(q;P_B) & := & e^{iP_BX} \int d^4(x-y) e^{iq(x-y)} \langle 0 | T 
 \psi(x) \overline{\psi}(y) | \Kmbf{P} \rangle,
 \label{AEQ:H091005:4} \\
 \overline{\chi}(q;P_B) & := & e^{-iP_BX} \int d^4(y-x) e^{iq(y-x)} \langle
 \Kmbf{P} | T \psi(y) \overline{\psi}(x) | 0 \rangle,
 \label{AEQ:H091005:4b}
\end{eqnarray}
with $X:=\frac{x+y}{2}$.
It should be noted that Eq.$(\ref{AEQ:H100301:1})$ is an exact
relation representing Eq.$(\ref{AEQ:H100318:4})$, where exact means
it holds for any $m_0$.

 The Ward--Takahashi
identity can be derived by the standard path integral procedure
\begin{eqnarray}
 \lefteqn{ -i \partial^\mu \langle 0 | T \psi(x) \overline{\psi}(y)
 \tilde{J}_{5\mu}^\alpha(0) | 0 \rangle = - 2i m_0
  \langle 0 | T \psi(x) \overline{\psi}(y) I^\alpha_5(0)
 | 0 \rangle }
  \label{AEQ:H091004:15} \\
& & {} + i\gamma_5 \frac{\lambda^\alpha}{2} \langle 0 | T \psi(x) 
\overline{\psi}(y) | 0 \rangle \delta^4(x) + \langle 0 | T \psi(x)
\overline{\psi}(y) | 0 \rangle i\gamma_5 \frac{\lambda^\alpha}{2} 
\delta^4(y). \nonumber 
\end{eqnarray}
Note that in Eq.$(\ref{AEQ:H091004:15})$ 
we replace the $T^*$ product to the $T$ product. 
(This proceduce is not rigorous but is justified
when $\tilde{J}^\alpha_{5\mu}(0)$ and $I^\alpha_5(0)$ are local currents.) 
 From this identity, one obtains the NG solution of
the BS amplitude in the limit of $m_0\to 0$,
\begin{equation}
 \chi(q;0) = \frac{1}{\tilde{f}_\pi} \{ i\frac{\lambda^\alpha}{2}\gamma_5,
  S_F(q) \}  \label{AEQ:H091005:7}
\end{equation}
where $S_F(q)$ is the full propagator of the quark.
 For a small quark mass $m_0$, one expects that 
Eq.$(\ref{AEQ:H091005:7})$ gives an
approximate solution of the BS equation.
An approximation of the exact relation Eq.$(\ref{AEQ:H100301:1})$ is
obtained by substituting Eq.$(\ref{AEQ:H091005:7})$ into 
Eq.$(\ref{AEQ:H100301:1})$ 
\begin{equation}
 M_\pi^2 \tilde{f}_\pi^2 \simeq -2m_0
 \langle \overline{\psi}\psi \rangle_0,  \quad 
 \langle \overline{\psi}\psi \rangle_0 := -\int_q f(-q^2)
 {\rm tr}[S_F(q)_{m_0=0}]. 
 \label{AEQ:H100301:3}
\end{equation}
The chiral condensate $\langle \overline{\psi}\psi \rangle_0$ is 
considered to be evaluated at the scale $\Lambda_{\rm UV}$.
To compare it the renormalized one in QCD, one must impose a
suitable renormalization condition.
This is the Gell-Mann--Oakes--Renner mass formula.

 In conclusion, we have formally proved that the low-energy relations
are satisfied if we redefine the axial-vector
current and the decay constant $\tilde{f}_\pi$ according
to the global chiral symmetry of the effective lagrangian.
 In the next section, we show how the above relations 
are described by the quark propagator and the BS amplitude
in the framework of the rainbow--ladder approximation.
 Later, we will see that the difference
between $f_\pi$ and $\tilde{f}_\pi$ is significantly large
in the improved ladder model.


\section{Analysis with CJT Action} \label{SEC:H100301:1}
 In this section, we consider the chiral symmetry property 
of the SD and BS equations when they are truncated. 
The similar discussion has been already done 
in the case of QED or for a specific  
interaction kernel which satisfies Eq.$(\ref{AEQ:H100308:8})$.\cite{HAR97,KA3}
However, it is non-trivial in the case of 
the general non-local interactions.
Especially it is difficult to find the formula of the decay constant
$\tilde{f}_\pi$ because the Noether current is complicated.
For the systematic 
discussion, we use the Cornwall, Jackiw and Tomboulis (CJT) 
effective action formulation.

 The Cornwall, Jackiw and Tomboulis effective action formulation
is one of the most powerful and useful methods to derive the SD and BS
equations consistently with the chiral symmetry.\cite{CJT,MIRANSKY,HAYMAKER}
 We study the chiral property
 with this formulation following Munczek.\cite{KA3}\, 
The CJT action is given by
\begin{equation}
 \Gamma[S_F] := i {\rm Tr}{\rm Ln}[S_F] - i{\rm Tr}[S_0^{-1}S_F] 
+ \Gamma_{\rm loop}[S_F].
 \label{AEQ:H091005:16}
\end{equation}
 The last term of Eq.$(\ref{AEQ:H091005:16})$, 
 $i\Gamma_{\rm loop}[S_F]$ is given by the sum of
 all Feynman amplitudes of 2-particle irreducible
 vacuum diagrams with two or more loops in  which every 
 bare quark propagator $S_0$
 is replaced by the full one $S_F$.
We will show that the truncation in the loop expansion preserves properties
of the chiral symmetry.

 Using the CJT action, the SD equation and the inhomogeneous BS equation are
derived by
\begin{equation}
 \frac{\delta \Gamma[S_F]}{\delta S_F(x,y)} = 0,
 \label{AEQ:H091005:18}
\end{equation}
\begin{equation}
 \frac{1}{i} \frac{\delta ^2 \Gamma[S_F]}{\delta S_{Fmn}(x,y) \delta
  S_{Fn'm'}(y',x')} G_{C;n'm'm''n''}^{(2)}(y'x';x''y'') =
   \delta_{m''m}\delta_{nn''}\delta(x''-x)\delta(y-y'')
 \label{AEQ:H091005:19}
\end{equation}
respectively,
where the repeated indices are summed and the repeated arguments
are  intergrated. $G_{C;n'm'm''n''}^{(2)}(y'x';x''y'')$ is the two-body 
connected Green function defined by
\begin{eqnarray}
 G_{C;nmm'n'}^{(2)}(yx;x'y') & := & \langle 0 | T \psi_n(y)
 \overline{\psi}_m(x)
 \psi_{m'}(x') \overline{\psi}_{n'}(y') | 0 \rangle  \nonumber \\
& & {} \quad - \langle 0 | T\psi_n(y)\overline{\psi}_m(x) |
 0 \rangle \langle 0 | T\psi_{m'}(x')\overline{\psi}_{n'}(y')
  | 0 \rangle. \label{AEQ:H091006:3}
\end{eqnarray}
By representing $G_{C;n'm'm''n''}^{(2)}(y'x';x''y'')$ in the spectral
form and taking the pion pole term, we express $G_C^{(2)}$ in terms of 
the BS amplitude of the pion as
\begin{eqnarray}
\lefteqn{ G^{(2)}_{C;nmm'n'}(yx;x'y')  =  \int_P  e^{-i(P-P_B)X+i(P-P_B)X'} } \nonumber \\
& & \times i\frac{\chi_{nm}(y,x;P_B)\overline{\chi}_{m'n'}(x',y';P_B)}
{P^2-M_\pi^2+i\epsilon} + R_{nmm'n'}(yx;x'y') \label{AEQ:H091006:4}
\end{eqnarray}
with
\begin{equation}
 X := \frac{ x + y}{2} ,\quad X' := \frac{ x' + y'}{2} ,
 \label{AEQ:H091006:5}
\end{equation}
\begin{equation}
 P=(P_0,\Kmbf{P}),\quad P_B=(\sqrt{M_\pi^2+\Kmbf{P}^2},\Kmbf{P}).
 \label{AEQ:H091006:7}
\end{equation}
The regular term $R_{nmm'n'}(yx;x'y')$ denotes the
contributions from excited states. 
The BS amplitude $\chi_{nm}(y,x;P_B)$ in Eq.$(\ref{AEQ:H091006:4})$ is a
solution of the homogeneous BS equation, given by
\begin{equation}
 \frac{\delta^2 \Gamma[S_F]}{\delta S_{Fmn}(x,y) \delta S_{Fn'm'}(y',x')}
  \chi_{n'm'}(y',x';P_B) = 0.
 \label{AEQ:H091006:8}
\end{equation}
In order to obtain the normalized BS amplitude, we go back to 
Eq.$(\ref{AEQ:H091005:19})$ and determine the normalization.

 The question is whether the chiral properties of the effective model
are preserved when 
various approximations are taken into account.
 To answer this let us consider a local infinitesimal
 axial transformation of $S_F(x,y)$
\begin{equation}
 S_F(x,y) \to S_F'(x,y) := (1+i\gamma_5 \frac{\lambda^\alpha}{2} \theta^\alpha(x)) S_F(x,y) (1+i\gamma_5 \frac{\lambda^\alpha}{2} \theta^\alpha(y)) \quad
 \label{AEQ:H091006:10}
\end{equation}
 Under this local transformation, the change of the CJT action is 
\begin{equation}
 \Delta_5 \Gamma[S_F] \equiv \frac{\delta \Gamma[S_F]}{\delta S_{Fn'm'}
 (y',x')}\{i\gamma_5 \frac{\lambda^\alpha}{2}\theta^\alpha,S_F\}_{n'm'
 }(y',x') \label{AEQ:H091109:2}
\end{equation}
with
\begin{equation}
 \{i\gamma_5 \frac{\lambda^\alpha}{2}\theta^\alpha,S_F\}(y,x) :=
 i\gamma_5 \frac{\lambda^\alpha}{2} \theta^\alpha(y) S_F(y,x) +
 S_F(y,x) i\gamma_5\frac{\lambda^\alpha}{2}\theta^\alpha(x).
 \label{AEQ:H100401:1}
\end{equation}
 Then the following equation holds:
\begin{eqnarray}
\lefteqn{ G^{(2)}_{C;m''n''nm}(x''y'';yx)\frac{\delta \Big( \Delta_5
\Gamma[S_F] \Big)}{\delta S_{Fnm}(y,x)}   } \nonumber \\
& \equiv & G^{(2)}_{C;m''n''nm}(x''y'';yx) \frac{\delta^2\Gamma[S_F]}{
\delta S_{Fnm}(y,x)\delta S_{Fm'n'}(x',y')}\{i\gamma_5\frac{\lambda^\alpha
}{2}\theta^\alpha,S_F\}_{m'n'}(x',y') \nonumber \\
& & {}+ \frac{\delta \Gamma[S_F]}{\delta S_{Fm'n'}(x',y')}\bigg\{ G^{(2)
}_{C;m''n''m'l}(x''y'';x'y')(i\gamma_5\frac{\lambda^\alpha}{2}
\theta^\alpha(y'))_{ln'} \nonumber \\
& & {} \quad +(i\gamma_5\frac{\lambda^\alpha}{2}\theta^\alpha(x')
)_{m'l}G^{(2)}_{C;m''n''ln'}(x''y'';x'y')\bigg\} \nonumber
 \\ 
& = & i\{i\gamma_5\frac{\lambda^\alpha}{2}\theta^\alpha,S_F\}_{m''n''}
(x'',y'').
 \label{AEQ:H091006:12}
\end{eqnarray}
 Here for the last equality we use the SD equation
$(\ref{AEQ:H091005:18})$ and
the inhomogeneous BS equation $(\ref{AEQ:H091005:19})$.
Eq.$(\ref{AEQ:H091006:12})$
is a key equation in our study of the system of the SD and BS equations.

 In the following, we consider three cases in which the local chiral
invariance is broken in different ways. First, we discuss the
naive case where neither cut-off nor momentum dependent
interactions are included. Second, we take into account the loop
cut-off function. Finally we consider the case where the interaction
kernel is not locally chiral invariant. In each case, we will show that the
"pion" becomes massless in the chiral limit, and present the formula
for the pion decay constant. We further show the PCAC relation in each
case.

\subsection{Naive Case} \label{SEC:H091224:2}
 First we consider the case that the cut-off function is
not introduced and the interaction does not modify the axial-vector current. 
 In this case, truncation to an arbitrary subset of diagrams in 
$\Gamma_{\rm loop}[S_F]$ does not violate the local chiral
invariance.\cite{KG5} Therefore, the following results are valid in 
each order of the loop expansion of $\Gamma_{\rm loop}[S_F]$. 
The first term $i{\rm Tr}{\rm Ln}[S_F]$ in Eq.$(\ref{AEQ:H091005:16})$
is also invariant because 
\begin{equation}
 \Delta_5 i{\rm Tr}{\rm Ln}[S_F]  \equiv 
 i{\rm Tr}[S_F^{-1} \{ i\gamma_5\frac{\lambda^\alpha}{2}\theta^\alpha,S_F\}]
  \equiv 2i{\rm tr}[i\gamma_5\frac{\lambda^\alpha}{2}]\int d^4x \theta^\alpha(x)
 \equiv 0. \label{AEQ:H100428:1}
\end{equation}
Then only the second term $-i{\rm Tr}[S_0^{-1}S_F]$
in Eq.$(\ref{AEQ:H091005:16})$ contributes in the left hand side(LHS) of
Eq.$(\ref{AEQ:H091006:12})$, and gives in momentum space
\begin{eqnarray}
\lefteqn{ \int_q G^{(2)}_{C;m''n''nm}(p,q;P)\Big( i\gamma_5 \frac{\lambda^\alpha}{2}(2m_0+\Slash{P})\Big)_{mn} } \nonumber \\
& = & i\Big( i\gamma_5\frac{\lambda^\alpha}{2}S_F(p-\frac{P}{2})\Big)_{m''n''}
 + i\Big( S_F(p+\frac{P}{2})i\gamma_5\frac{\lambda^\alpha}{2}\Big)_{m''n''}.
 \label{AEQ:H091007:4}
\end{eqnarray}
 Here the Fourier transformation of $G^{(2)}_{C;m'n'nm}(x'y';yx)$
is defined by
\begin{equation}
 G^{(2)}_{C;m'n'nm}(x'y';yx) = \int_{pqP}e^{-i\{p(x'-y')+q(y-x)+P(X'-X)\}}
 G^{(2)}_{C;m'n'nm}(p,q;P).
 \label{AEQ:H091011:10}
\end{equation}
 Using Eq.$(\ref{AEQ:H091006:4})$, Eq.$(\ref{AEQ:H091007:4})$ becomes
\begin{eqnarray}
 \lefteqn{ \int_q \bigg[ i \frac{\chi_{m''n''}(p;P_B)\overline{\chi}_{nm}
(q;P_B)}{P^2-M_\pi^2+i\epsilon}+R_{m''n''nm}(p,q;P) \bigg] \Big( i\gamma_5\frac{\lambda^\alpha}{2}(2m_0+\Slash{P})\Big)_{mn} } \nonumber \\
 & = & i\Big(i\gamma_5\frac{\lambda^\alpha}{2}S_F(p-\frac{P}{2})\Big)_{m''n''}+i\Big(S_F(p+\frac{P}{2})i\gamma_5\frac{\lambda^\alpha}{2}\Big)_{m''n''}
 \label{AEQ:H091007:5}
\end{eqnarray}
with
\begin{eqnarray}
 \chi_{mn}(x,y;P_B) & = & e^{-iP_BX}\int_q e^{-iq(x-y)}
 \chi_{mn}(q;P_B),  \label{AEQ:H091018:2} \\
 \overline{\chi}_{nm}(y,x;P_B) & = & e^{iP_BX}\int_q 
 e^{-iq(y-x)}\overline{\chi}_{nm}(q;P_B). \label{AEQ:H091018:1} 
\end{eqnarray}

\subsubsection{Chiral Limit}
 In the chiral limit $m_0\to 0$, the CJT action is chiral invariant.
We confirm that the "pion" becomes massless in this limit because if
$M_\pi\ne 0$ LHS of Eq.$(\ref{AEQ:H091007:5})$ diverges
in the limit $P^2\to M_\pi^2$, while RHS
converges.
In this case LHS of Eq.$(\ref{AEQ:H091007:5})$ becomes
\begin{eqnarray}
 \lefteqn{ \mbox{LHS of Eq.}(\ref{AEQ:H091007:5}) = \int_q \frac{i}{P^2}{\rm tr}[\overline{\chi}(q;P_B)i\gamma_5\frac{\lambda^\alpha}{2}\Slash{P}]\chi_{m''n''}(p;P_B) } \nonumber \\
 & & {} + \int_q R_{m''n''nm}(p,q;P)(i\gamma_5\frac{\lambda^\alpha}{2}\Slash{P})_{mn}. \label{AEQ:H091007:6}
\end{eqnarray}
 If we take the soft limit $\Kmbf{P}\to 0$ after taking the on-shell
limit $P\to P_B$ (i.e. $P_0\to \sqrt{\Kmbf{P}^2}$), the second term of
Eq.$(\ref{AEQ:H091007:6})$ vanishes and the first term becomes
\begin{equation}
 \mbox{1st term of Eq.}(\ref{AEQ:H091007:6}) \to i f_\pi \chi_{m''n''}(p;0),
 \label{AEQ:H091007:7}
\end{equation}
with
\begin{equation}
 f_\pi = \lim_{P\to P_B} \frac{1}{P^2} \int_q 
{\rm tr}[\overline{\chi}(q;P_B) i\gamma_5\frac{\lambda^\alpha}{2}\Slash{P}].
 \label{AEQ:H091007:8}
\end{equation}
 On the other hand RHS of Eq.$(\ref{AEQ:H091007:5})$ becomes
\begin{equation}
 \mbox{RHS of Eq.}(\ref{AEQ:H091007:5}) \to i\{i\gamma_5\frac{\lambda^\alpha}{2},S_F(p)\}_{m''n''}
\end{equation}
in the same limit. Then we obtain the equation
\begin{equation}
 \chi_{mn}(q;0) = \frac{1}{f_\pi} \{i\gamma_5 \frac{\lambda^\alpha}{2},S_F(q)\}_{mn}.
 \label{AEQ:H091007:9}
\end{equation}
 This is the same equation as Eq.$(\ref{AEQ:H091005:7})$ which was derived
from the axial-vector Ward--Takahashi identity.
It should be noted that Eq.$(\ref{AEQ:H091007:9})$ is
valid in the truncated SD and 
BS equations in which $\Gamma_{\rm loop}[S_F]$ is expanded to a finite
number of loops.

\subsubsection{Finite Quark Mass}
 When $m_0>0$, the first term in the brackets of Eq.$(\ref{AEQ:H091007:5})$
will diverge in the on-shell limit $P\to P_B$. Then the equation
\begin{equation}
 \int_q {\rm tr}[\overline{\chi}(q;P_B)i\gamma_5\frac{\lambda^\alpha}{2}(2m_0+\Slash{P}_{\!B} )] = 0 \label{AEQ:H091007:10} 
\end{equation}
must be satisfied and we obtain
\begin{equation}
 M_\pi^2 f_\pi = -2m_0 i \int_q {\rm tr}[\overline{\chi}(q;P_B)\gamma_5\frac{\lambda^\alpha}{2}]
 \label{AEQ:H091007:11}
\end{equation}
where $f_\pi$ is given by Eq.$(\ref{AEQ:H091007:8})$.
This equation is again same as Eq.$(\ref{AEQ:H100301:1})$.
Using the approximation Eq.$(\ref{AEQ:H091007:9})$, we obtain the GMOR
relation 
\begin{equation}
 M_\pi^2 f_\pi^2 \simeq -2m_0 \langle \overline{\psi} \psi \rangle_0 ,\quad
\langle \overline{\psi}\psi \rangle_0 := - \int_q {\rm tr}[S_F(q)_{m_0=0}]. \label{AEQ:H091007:12}
\end{equation}
 Note again that any truncation of $\Gamma_{\rm loop}[S_F]$ 
is guaranteed to satisfy all the above equations.

\subsection{The Case of the Cut-off Regularization} \label{SSEC:H091213:3}
 In order to regularize loop integrals, we introduce the cut-off
function in the kinetic term of the lagrangian as 
in Eq.$(\ref{AEQ:H100308:2})$.
As far as this regularization is taken, the interaction term of 
the effective action is not modified and therefore the chiral invariance of 
$\Gamma_{\rm loop}[S_F]$ is not violated.
 When the cut-off function is introduced, the same procedure can be applied
 with the replacement  
\begin{eqnarray}
 \lefteqn{
  i\gamma_5\frac{\lambda^\alpha}{2}(2m_0 + \Slash{P}) } 
   \label{AEQ:H091011:5} \\
 & \mapsto & i\gamma_5\frac{\lambda^\alpha}{2} \left( 2m_0 
 \frac{f(-q_-^2)+f(-q_+^2)}{2} - f(-q_-^2)\Slash{q}_-+f(-q_+^2)\Slash{q}_+
  \right), \label{AEQ:H091011:6}
\end{eqnarray}
\begin{equation}
 q_- := q-\frac{P}{2},\quad q_+ := q+\frac{P}{2}. \label{AEQ:H091011:7}
\end{equation}
As a result the decay constant $\tilde{f}_\pi$ is given by
\begin{equation}
 \tilde{f}_\pi = \lim_{P\to P_B} \frac{1}{P^2} \int_q {\rm tr}\bigg[
 \overline{\chi}(q;P_B) i\gamma_5\frac{\lambda^\alpha}{2} \bigg\{
 \frac{f(-q_-^2)+f(-q_+^2)}{2}\Slash{P} + (f(-q_+^2)-f(-q_-^2))
 \Slash{q} \bigg\} \bigg]. \label{AEQ:H091011:9}
\end{equation}
This $\tilde{f}_\pi$ coincides with the definition $(\ref{AEQ:H100225:1})$.
 Therefore the exact relation Eq.$(\ref{AEQ:H091007:11})$ is modified to
\begin{equation}
 M_\pi^2 \tilde{f}_\pi = -2m_0 i \int_q  \frac{f(-q_{-B}^2)+f(-q_{+B}^2)}{2}
 {\rm tr} [\overline{\chi}(q;P_B)\gamma_5\frac{\lambda^\alpha}{2}]
 \label{AEQ:H091011:8}
\end{equation}
and the definition of quark condensate is modified to
\begin{equation}
 \langle \overline{\psi}\psi \rangle_0 = -\int_q f(-q^2) {\rm tr}[S_F(q)_{m_0=0}]. \label{AEQ:H091013:7}
\end{equation}
 Under these modifications, Eqs.$(\ref{AEQ:H091007:9})$ and $(\ref{AEQ:H091007:12})$ hold.

\subsection{The Case of the Non-local Interaction} \label{SSEC:H091213:6}
 Here we consider the case that the interaction modifies the axial-vector
current.
To simplify the argument, we omit the cut-off function $f(\zeta)$ in 
this subsection.
 We employ the two-loop approximation of
$\Gamma_{\rm loop}[S_F]$ such that
\begin{eqnarray}
 \Gamma_{\rm loop}[S_F] & = & -\frac{1}{2}\int d^4 x {\cal K}^{m_1m_2,n_1n_2}
 \left( i\partial_{x_1},i\partial_{x_2};i\partial_{y_1},i\partial_{y_2}\right)
  \label{AEQ:H091007:15b}\\
 & & {} \times \left[ S_{Fm_2m_1}(x_2,x_1)S_{Fn_2n_1}(y_2,y_1)
 -S_{Fm_2n_1}(x_2,y_1)S_{Fn_2m_1}(y_2,x_1)\right]\big|_* \nonumber 
\end{eqnarray}
where the symbol $*$ means to take $x_1,x_2,y_1,y_2\to x$ after all the
derivatives are operated.
 Note that this approximation corresponds to the rainbow approximation
in the SD equation and the ladder approximation in the BS equation.
Then Eq.$(\ref{AEQ:H091006:12})$ reduces in momentum space to
\begin{eqnarray}
\lefteqn{ \int_q G^{(2)}_{C;m''n''nm}(p,q;P)\Big( i\gamma_5 
\frac{\lambda^\alpha}{2}(2m_0+\Slash{P})+E^\alpha(q;P)\Big)_{mn} } 
\nonumber \\
& = & i\Big( i\gamma_5\frac{\lambda^\alpha}{2}S_F(p-\frac{P}{2})
\Big)_{m''n''} + i\Big( S_F(p+\frac{P}{2})i\gamma_5
\frac{\lambda^\alpha}{2}\Big)_{m''n''},
 \label{AEQ:H091007:13}
\end{eqnarray}
\begin{eqnarray}
 \lefteqn{ E^\alpha_{mn}(q;P) := \int_k \bigg[
  \bigg\{  {\cal K}^{ln,n'm'} (-q-\frac{P}{2},q+\frac{P}{2};-k,k) } 
  \nonumber \\
  & & {}\quad
    - {\cal K}^{ln,n'm'}(-q+\frac{P}{2},q-\frac{P}{2};-k,k) \bigg\}
  (i\gamma_5\frac{\lambda^\alpha}{2})_{ml} S_{Fm'n'}(k)  \nonumber \\
 & & 
 {} + \bigg\{ {\cal K}^{n'm',mn}(-k+P,k;-q-\frac{P}{2},q-\frac{P}{2}) 
 \nonumber \\
 & & {} \quad
 - {\cal K}^{n'm',mn}(-k,k+P;-q-\frac{P}{2},q-\frac{P}{2}) \bigg\}
 (i\gamma_5\frac{\lambda^\alpha}{2}S_F)_{m'n'}(k) \nonumber \\
 & & 
 {} + \bigg\{ {\cal K}^{n'n,mm'}(-k,q-\frac{P}{2};-q-\frac{P}{2},k+P)
 \nonumber \\
 & & {} \quad
 - {\cal K}^{n'n,mm'}(-k,q-\frac{P}{2};-q+\frac{P}{2},k) \bigg\}
 (i\gamma_5\frac{\lambda^\alpha}{2}S_F)_{m'n'}(k) \nonumber \\
 & & 
 {} + \bigg\{ {\cal K}^{n'n,mm'}(-k+P,q-\frac{P}{2};-q-\frac{P}{2},k)
 \nonumber \\
 & & {} \quad
 - {\cal K}^{n'n,mm'}(-k,q+\frac{P}{2};-q-\frac{P}{2},k) \bigg\}
 (S_Fi\gamma_5\frac{\lambda^\alpha}{2})_{m'n'}(k) \bigg].
 \label{AEQ:H091007:14}
\end{eqnarray}
 Substituting Eq.$(\ref{AEQ:H091006:4})$, Eq.$(\ref{AEQ:H091007:13})$ becomes
\begin{eqnarray}
 \lefteqn{ \int_q \bigg[ i \frac{\chi_{m''n''}(p;P_B)\overline{\chi}_{nm}(q;P_B)}{P^2-M_\pi^2+i\epsilon}+R_{m''n''nm}(p,q;P) \bigg] \Big( i\gamma_5\frac{\lambda^\alpha}{2}(2m_0+\Slash{P})+E^\alpha(q;P)\Big)_{mn} } \nonumber \\
 & = & i\Big(i\gamma_5\frac{\lambda^\alpha}{2}S_F(p-\frac{P}{2})\Big)_{m''n''}+i\Big(S_F(p+\frac{P}{2})i\gamma_5\frac{\lambda^\alpha}{2}\Big)_{m''n''}.
 \label{AEQ:H091007:15}
\end{eqnarray}
It is easy to see that $E^\alpha(q;P)$ becomes zero
in the limit $P\to 0$. This is expected
because LHS of Eq.$(\ref{AEQ:H091007:13})$ or
Eq.$(\ref{AEQ:H091007:15})$ should be finite in
the chiral limit as RHS is.
It is also easy to check that $E^\alpha(q;P)$ vanishes if 
${\cal K}^{mm',nn'}(p,p';q,q')$
satisfies Eq.$(\ref{AEQ:H100308:8})$.

\subsubsection{Chiral Limit}
 In the chiral limit, the global chiral symmetry breaking implies
the existence of the NG boson again. Then we obtain
\begin{eqnarray}
 \lefteqn{ \mbox{LHS of Eq.}(\ref{AEQ:H091007:15}) = \int_q \frac{i}{P^2}{\rm tr}[\overline{\chi}(q;P_B)(i\gamma_5\frac{\lambda^\alpha}{2}\Slash{P}+E^\alpha(q;P))]\chi_{m''n''}(p;P_B) } \nonumber \\
 & & {} + \int_q R_{m''n''nm}(p,q;P)(i\gamma_5\frac{\lambda^\alpha}{2}\Slash{P})_{mn}. \label{AEQ:H091007:16}
\end{eqnarray}
 Taking the soft limit $\Kmbf{P}\to 0$ after the on-shell limit
$P\to P_B$ (i.e. $P_0\to \sqrt{\Kmbf{P}^2}$) the second term of
Eq.$(\ref{AEQ:H091007:16})$ vanishes and the first term becomes
\begin{equation}
 \mbox{1st term of Eq.}(\ref{AEQ:H091007:16}) \to i \tilde{f}_\pi \chi_{m''n''}(p;0)
 \label{AEQ:H091007:17}
\end{equation}
where we use a (new) definition of $\tilde{f}_\pi$,
\begin{equation}
 \tilde{f}_\pi := \lim_{P\to P_B} \frac{1}{P^2} \int_q
 {\rm tr}[\overline{\chi}(q;P_B) (i\gamma_5\frac{\lambda^\alpha}{2}
\Slash{P}+E^\alpha(q;P))]. \label{AEQ:H091007:18}
\end{equation}

 On the other hand RHS of Eq.$(\ref{AEQ:H091007:15})$ becomes
\begin{equation}
 \mbox{RHS of Eq.}(\ref{AEQ:H091007:15}) \to i\{i\gamma_5\frac{\lambda^\alpha}{2},S_F(p)\}_{m''n''}
\end{equation}
in the same limit. Thus we obtain the relation
\begin{equation}
 \chi_{mn}(q;0) = \frac{1}{\tilde{f}_\pi} \{i\gamma_5
 \frac{\lambda^\alpha}{2},S_F(q)\}_{mn}.
 \label{AEQ:H091007:19}
\end{equation}
 This is the same equation as Eq.$(\ref{AEQ:H091005:7})$ which was derived
from the axial-vector Ward--Takahashi identity.
 Therefore we conclude that the definition
of $\tilde{f}_\pi$ in Eq.$(\ref{AEQ:H091007:18})$ is equivalent to the
definition Eq.$(\ref{AEQ:H100225:1})$.

\subsubsection{Finite Quark Mass}
 When $m_0>0$, the first term in the brackets of Eq.$(\ref{AEQ:H091007:15})$ diverges in the on-shell limit $P\to P_B$. Then the equation
\begin{equation}
 \int_q {\rm tr}[\overline{\chi}(q;P_B)(i\gamma_5\frac{\lambda^\alpha}{2}(2m_0+\Slash{P}_B)+E^\alpha(q;P_B)) ] = 0 \label{AEQ:H091007:20} 
\end{equation}
must hold. From Eq.$(\ref{AEQ:H091007:18})$ we obtain
\begin{equation}
 M_\pi^2 \tilde{f}_\pi = -2m_0 i \int_q {\rm tr}[\overline{\chi}(q;P_B)\gamma_5\frac{\lambda^\alpha}{2}].
 \label{AEQ:H091007:21}
\end{equation}
This equation again coincides with Eq.$(\ref{AEQ:H100301:1})$
when $f(\zeta)\equiv 1$ and
implies that our choice of
$\tilde{f_\pi}$ in Eq.$(\ref{AEQ:H091007:18})$ is consistent with
Eq.$(\ref{AEQ:H100225:1})$. The GMOR relation is
written as
\begin{equation}
 M_\pi^2 \tilde{f}_\pi^2 \simeq -2m_0 \langle \overline{\psi} \psi \rangle_0 ,\quad
\langle \overline{\psi}\psi \rangle_0 := - \int_q {\rm tr}[S_F(q)_{m_0=0}]. \label{AEQ:H091007:22}
\end{equation}
Thus we have proved that the truncation of $\Gamma_{\rm loop}[S_F]$
preserves the low energy property of the effective model if one uses
the appropriate formula of the decay constant given by
Eq.$(\ref{AEQ:H091007:18})$.

 Up to now, we neglect the cut-off function.
 Introducing both the effects of the cut-off function 
and the non-local interaction,
the formula for the decay constant in the approximation
 Eq.$(\ref{AEQ:H091007:15b})$ is given by
\begin{eqnarray}
 \tilde{f}_\pi & = & \lim_{P\to P_B} \frac{1}{P^2} \int_q {\rm tr}\bigg[
 \overline{\chi}(q;P_B) \bigg\{ i\gamma_5\frac{\lambda^\alpha}{2} \bigg(
 \frac{f(-q_-^2)+f(-q_+^2)}{2}\Slash{P} + (f(-q_+^2)-f(-q_-^2))
 \Slash{q} \bigg) \nonumber \\
 & & {} \quad + E^\alpha(q;P) \bigg\} \bigg].  \label{AEQ:H100308:21}
\end{eqnarray}
It should be noted here that we have defined $\tilde{f}_\pi$ 
in Eq.$(\ref{AEQ:H091007:18})$ so as to reproduce the
low-energy relations Eqs.$(\ref{AEQ:H091007:19})$ and $(\ref{AEQ:H091007:21})$.
The definition Eq.$(\ref{AEQ:H091007:18})$ is obtained 
from Eq.$(\ref{AEQ:H100225:1})$ directly by taking the
following approximations
\begin{eqnarray}
 \langle 0 | T \psi \overline{\psi} | \Kmbf{P} \rangle 
& \mapsto & \langle 0 | T \psi \overline{\psi} | \Kmbf{P} 
\rangle_{\rm ladder}, 
\label{AEQ:H100428:2} \\
\langle 0 | T \psi \overline{\psi} \psi \overline{\psi} | \Kmbf{P} \rangle 
& \mapsto & \sum \langle 0 | T \psi \overline{\psi} | 0 \rangle_{\rm rainbow}
\langle 0 | T \psi \overline{\psi} | \Kmbf{P} \rangle_{\rm ladder}.
\label{AEQ:H100428:3}
\end{eqnarray}


\section{General Formula of the Decay Constant} \label{SEC:H100409:1}
 So far we have treated the lowest order (rainbow-ladder) approximation
in the CJT action.
When we proceed to the higher order terms, the two-loop formula
Eq.$(\ref{AEQ:H100308:21})$ should be modified to an appropriate form.
 Then we can prove that the truncation of $\Gamma_{\rm loop}[S_F]$
preserves 
the low energy property of the effective model.
We note that the consistency is guaranteed in the loop expansion of the
CJT action formulation. Further approximations
inconsistent with the loop expansion will 
violate the low energy properties.
An example is to take the
leading terms of the Chebychev polynomial expansion of the BS amplitude,
although the numerical result shows
that the violation of the PCAC relation is generally small.\cite{BQRT,MR}

 In order to derive a general formula of the decay constant, we consider
the transformation property of the CJT action.
 Under the infinitesimal local axial transformation, the change 
of the classical action can be written  as
\begin{equation}
 \Delta_5 S[\psi,\overline{\psi}] = \int d^4x \theta^\alpha(x)
  \Big(\partial^\mu \tilde{J}^\alpha_{5\mu}[\psi,\overline{\psi}](x) -
   M^\alpha[\psi,\overline{\psi}](x) \Big).
 \label{AEQ:H091011:2}
\end{equation}
 Here $M^\alpha[\psi,\overline{\psi}](x)$ comes from the globally variant
terms, such as the quark mass term and
$\tilde{J}^\alpha_{5\mu}[\psi,\overline{\psi}](x)$
is an effective axial-vector current.

 The change of the CJT action corresponding to the classical action must
be written  as 
\begin{equation}
 \Delta_5 \Gamma[S_F] = \int d^4 x \theta^\alpha(x) \Big( \partial^\mu
 \tilde{\cal J}^\alpha_{5\mu}[S_F](x) - {\cal M}^\alpha[S_F](x) \Big).
 \label{AEQ:H091011:3}
\end{equation}
 Again ${\cal M}^\alpha[S_F](x)$ comes from the globally
variant terms and $\tilde{\cal J}^\alpha_{5\mu}[S_F](x)$ is a proxy of the
effective axial-vector current in the CJT action. 

 When the effective axial-vector current
$\tilde{J}^\alpha_{5\mu}[\psi,\overline{\psi}](x)$ has a non-local
interaction term, the exact proxy $\tilde{\cal J}^\alpha_{5\mu}[S_F](x)$ is
an infinite sum of the Feynman amplitudes which come from the 
expansion of $\Gamma_{\rm loop}[S_F]$.
If one truncates the expansion of $\Gamma_{\rm loop}[S_F]$, the approximated 
proxy $\tilde{\cal J}^\alpha_{5\mu}[S_F](x)$ is built of a finite sum of the
Feynman amplitudes.

 For example, in the lowest loop approximation Eq.$(\ref{AEQ:H091007:15b})$
and with the cut-off function $f(\zeta)$ we obtain
\begin{eqnarray}
\lefteqn{
 \partial^\mu \tilde{{\cal J}}^\alpha_{5\mu}[S_F](z) = \int d^4x \int_p e^{-ip(z-x)}
 f(-p^2) \bigg\{ {\rm tr}[\Slash{p}i\gamma_5\frac{\lambda^\alpha}{2} S_F(z,x)]
- {\rm tr}[\Slash{p}S_F(x,z)i\gamma_5\frac{\lambda^\alpha}{2} ] \bigg\}
 } \nonumber \\
& & {} - \int_k e^{ikz} \int d^4x  {\cal K}^{m_1m_2,n_1n_2}\left( 
 i\partial_{x_1},i\partial_{x_2};i\partial_{y_1},i\partial_{y_2} \right)
 \bigg[ \nonumber \\
 & & \quad S_{Fm_2m_1}(x_2,x_1) \bigg\{ (S_Fi\gamma_5\frac{\lambda^\alpha}{2})_{n_2n_1}(y_2,y_1) e^{-iky_1} 
 + (i\gamma_5\frac{\lambda^\alpha}{2} S_F)_{n_2n_1}(y_2,y_1) e^{-iky_2}
  \bigg\}  \label{AEQ:H091223:9} \\
 & & {} - S_{Fm_2n_1}(x_2,y_1) \bigg\{ (S_Fi\gamma_5\frac{
 \lambda^\alpha}{2})_{n_2m_1}(y_2,x_1)e^{-ikx_1} + (i\gamma_5\frac{
 \lambda^\alpha}{2}S_F)_{n_2m_1}(y_2,x_1)e^{-iky_2} \bigg\} \bigg]
  \bigg|_*\,,\nonumber 
\end{eqnarray}
\begin{equation}
 {\cal M}^\alpha[S_F](z) = -m_0 \int d^4 x \int_p e^{-ip(z-x)}
  f(-p^2) \bigg\{
  {\rm tr}[i\gamma_5\frac{\lambda^\alpha}{2}S_F(z,x)] +
  {\rm tr}[S_F(x,z)i\gamma_5\frac{\lambda^\alpha}{2}] \bigg\}.
  \label{AEQ:H091223:10}
\end{equation}

 To calculate the decay constant, we define the formula
\begin{eqnarray}
 \tilde{f}_\pi & = & \lim_{P\to P_B} \frac{iP^\mu}{P^2}\overline{\chi}_{nm}(x,y;P_B) 
 \frac{\delta \tilde{{\cal J}}^\alpha_{5\mu}[S_F](0)}{\delta S_{Fnm}(x,y)} 
 \label{AEQ:H091011:4} \\
 & = & \lim_{P\to P_B} \frac{1}{P^2}\overline{\chi}_{nm}(x,y;P_B) 
 \frac{\delta  \partial^\mu  \tilde{{\cal J}}^\alpha_{5\mu}[S_F](0)}{\delta
  S_{Fnm}(x,y)}. \label{AEQ:H091223:11}
\end{eqnarray}
corresponding to the exact definition $(\ref{AEQ:H100225:1})$, or  
\begin{equation}
 \tilde{f}_\pi := \lim_{P\to P_B} \frac{iP^\mu}{P^2} \langle \Kmbf{P}
  |  \tilde{J}^\alpha_{5\mu}[\psi,\overline{\psi}](0) | 0 \rangle.
 \label{AEQ:H100308:20}
\end{equation}
 Therefore systematically approximated $\tilde{f}_\pi$ can be obtained for
any truncation of $\Gamma_{\rm loop}[S_F]$.
The previous formula Eq.$(\ref{AEQ:H100308:21})$ coincides with 
Eq.$(\ref{AEQ:H091223:11})$ if the rainbow-ladder
approximation is employed.

\subsection{Pagels--Stokar Formula} \label{SUBSEC:H100326:2}
 Pagels and Stokar proposed a useful approximation for
 the decay constant $f_\pi$ in the chiral limit 
in terms of the constituent quark mass function $B(q^2)$ of the 
SD equation.\cite{MIRANSKY,PS}
The Pagels--Stokar formula is
\begin{equation}
 f_\pi^2 = \frac{N_C}{8\pi^2} \int_{0}^{\infty} dq_E^2 q_E^2 
 \frac{B(-q_E^2)\Big[ 2B(-q_E^2) + q_E^2 B'(-q_E^2) \Big]}{(q_E^2+
 B^2(-q_E^2))^2} 
 \label{AEQ:H100202:2}
\end{equation}
where $B(-q_E^2)$ is defined by the ansatz 
\begin{equation}
 S_F(q) = \frac{i}{\Slash{q}-B(q^2)} 
 \label{AEQ:H100202:3}
\end{equation}
instead of the general form of the SD solution
\begin{equation}
 S_F(q) = \frac{i}{A(q^2)\Slash{q}-B(q^2)}.
 \label{AEQ:H100308:22}
\end{equation}
$B'(x)$ denotes the derivative of $B(x)$
and $q_E$ denotes the Euclidean momentum i.e. $q_E^2=-q^2$. 

 This formula Eq.$(\ref{AEQ:H100202:2})$ can be derived from 
an approximated BS amplitude
with the Ward--Takahashi identity for the axial-vector current. 
Therefore when 
the axial-vector current is modified, 
it should also be modified. In this section, we propose 
a new formula similar to the Pagels--Stokar
formula in the effective model.

 The BS amplitude is given by
\begin{equation}
 \chi_{nm}(q;0) = \frac{1}{\tilde{f}_\pi} \{i\gamma_5
 \frac{\lambda^\alpha}{2},S_F(q)\}_{nm} \label{AEQ:H091219:1}
\end{equation}
from the Ward--Takahashi identity for the axial-vector current
 in the chiral limit. 
But this solution is not sufficient to calculate $\tilde{f}_\pi$, because
$\tilde{f}_\pi$ is related to the derivative of the BS amplitude
with respect to the total momentum $P_B := (\sqrt{\Kmbf{P}^2},\Kmbf{P})$.
 The amputated BS amplitude(or BS vertex) $\hat{\chi}_{nm}(q;P_B)$
is defined by
\begin{eqnarray}
 \hat{\chi}_{nm}(q;P_B) := S^{-1}_{Fnn'}(q+\frac{P_B}{2}) \chi_{n'm'}(q;P_B) 
 S^{-1}_{Fm'm}(q-\frac{P_B}{2}), \label{AEQ:H091219:2}
\end{eqnarray}
and in the chiral limit, we obtain from Eq.$(\ref{AEQ:H091219:1})$
\begin{equation}
 \hat{\chi}_{nm}(q;0) = \frac{1}{\tilde{f}_\pi} \{i\gamma_5
  \frac{\lambda^\alpha}{2},S^{-1}_F(q)\}_{nm}. \label{AEQ:H091219:3}
\end{equation}
 Consider an approximation for the BS amplitude
\begin{eqnarray}
 \chi^{\rm PS}_{nm}(q;P_B) &= &  S_{Fnn'}(q+\frac{P_B}{2})
  \hat{\chi}_{n'm'}(q;0) S_{Fm'm}(q-\frac{P_B}{2})
  \label{AEQ:H091220:1} \\
 & = & \frac{1}{\tilde{f}_\pi} S_{Fnn'}(q+\frac{P_B}{2})
 \{ i\gamma_5\frac{\lambda^\alpha}{2},S^{-1}_F(q)\}_{n'm'}
 S_{Fm'm}(q-\frac{P_B}{2}).
 \label{AEQ:H091220:2}
\end{eqnarray}
In this paper we call this the Pagels--Stokar ansatz.
This is a very useful formula because it gives the approximated BS amplitude
for the "pion" in terms only of the solution of the SD equation.
As this specifies the dependence on the total momentum $P_B$, 
one can estimate the decay constant
without solving the BS equation, for instance 
using our formula Eq.$(\ref{AEQ:H100308:21})$ or Eq.$(\ref{AEQ:H091223:11})$.
The result is given by
\begin{eqnarray}
\lefteqn{
 (\tilde{f}^{\rm PS}_\pi)^2 = \lim_{P\to P_B} \frac{1}{P^2} \int_q
{\rm tr}\bigg[ S_F(q-\frac{P_B}{2})\{i\gamma_5\frac{
\lambda^\alpha}{2},
S^{-1}_F(q)\}S_F(q+\frac{P_B}{2}) }\nonumber \\
& & \times \bigg\{ i\gamma_5\frac{\lambda^\alpha}{2} \bigg(
 \frac{f(-q_-^2)+f(-q_+^2)}{2}\Slash{P} + (f(-q_+^2)-f(-q_-^2))
 \Slash{q} \bigg) + E^\alpha(q;P) \bigg\} \bigg]
 \label{AEQ:H100308:23}
\end{eqnarray}
where no summation with respect to $\alpha$ is taken.
This formula Eq.$(\ref{AEQ:H100308:23})$ is 
useful to estimate $\tilde{f}_\pi$ while the validity of the
Pagels--Stokar ansatz Eq.$(\ref{AEQ:H091220:2})$ is not established.
 Eq.$(\ref{AEQ:H100308:23})$
reduces to the original Pagels--Stokar formula $(\ref{AEQ:H100202:2})$
after the Wick rotation 
when (i) one neglects the effects of the
cut-off function $f(\zeta)$ and the local 4-quark interaction
to the axial-vector current, and (ii) if one assumes the ansatz  
$(\ref{AEQ:H100202:3})$ of the SD solution.


\section{Numerical Results for a Concrete Example} \label{SEC:H100326:3}
 In this section we show some numerical results as an example of our
approach. We here employ the effective model introduced 
by Aoki et al.\cite{KG1}.
The interaction is given by the one gluon exchange with the
running coupling constant which depends on the special set of
momenta according to the Higashijima--Miransky approximation. 
In the infrared region the running coupling constant is assumed to
be constant.  
We do not elucidate the detail of this model here.
(See Ref.\cite{TIT01}.)
Our choice of the parameters is
$\Lambda_{\rm QCD}=500$MeV, $t_{\rm IF}=-0.5$, $t_0=-3.0$ and the 
ultraviolet cut-off paramater $\Lambda_{\rm UV}=2.0$GeV.
We have solved the SD equation and the BS equation for the pion in the
Euclidean momentum in the chiral limit. As is pointed out in Ref.\cite{KG1},
the consistency of the SD and BS equations in the CJT action formulation
guarantees that the pion becomes massless in the chiral limit.
We have confirmed this in our numerical calculation.

In the rainbow--ladder approximation, the exact value of the pion
decay constant in the chiral limit is given by
Eq.$(\ref{AEQ:H100308:21})$ with $P_B^2=0$, which gives
\begin{equation}
 \tilde{f}_\pi = 72 {\rm MeV}.
 \label{AEQ:H100321:1}
\end{equation}
This can be first compared with the naive value $f_\pi$ in which
$E^\alpha(q,p)$ term coming from the non-local interaction is
omitted from Eq.$(\ref{AEQ:H100308:21})$ (See Eq.$(\ref{AEQ:H091007:8})$).
The result\footnotemark is
\begin{equation}
 f_\pi = 122 {\rm MeV}.
\label{AEQ:H100321:2}
\end{equation}
\footnotetext{ The cut-off regularization is applied for 
the naive value Eq.$(\ref{AEQ:H100321:2})$. } 
This is more than $70\%$ deviated from the true value.
Thus we see that the correction term $E^\alpha(q,p)$ is essential. 
In Ref.\cite{KG1}, 
the authors normalize the BS amplitude by
the true decay constant $\tilde{f}_\pi$ as in Eq.$(\ref{AEQ:H091219:1})$.
To calculate the decay constant, however, the naive definition 
Eq.$(\ref{AEQ:H100301:2})$ is employed.
As a result, their definition of the decay constant gives
\begin{equation}
  \sqrt{ f_\pi \tilde{f}_\pi } = 94 {\rm MeV}.
 \label{AEQ:H100321:3}
\end{equation}
 Next we examine the Pagel--Stokar formula.
The original
Pagels--Stokar formula Eq.$(\ref{AEQ:H100202:2})$ gives
\begin{equation}
 f_\pi^{\rm PS} = 96 {\rm MeV}
 \label{AEQ:H100321:4}
\end{equation}
which is close to Eq.$(\ref{AEQ:H100321:3})$ as is pointed out\footnotemark 
in Ref.\cite{KG1}.
\footnotetext{
The value of $f_\pi^{\rm PS}$ is different from Ref.\cite{KG1},
because of the different choice of the parameters 
$t_{\rm IF},\Lambda_{\rm QCD}$
and $\Lambda_{\rm UV}$.}
Our new formula Eq.$(\ref{AEQ:H100308:23})$ gives
\begin{equation}
 \tilde{f}_\pi^{\rm PS} = 73 {\rm MeV}
\end{equation}
which almost coincides with Eq.$(\ref{AEQ:H100321:1})$.
Thus we find that the new formula gives rather good prediction, although
only the SD equation is to be solved in order to calculate 
Eq.$(\ref{AEQ:H100308:23})$. Thus this is the most economical way to
estimate the pion decay constant.


\section{Conclusion} \label{SEC:H100327:1}
 In this paper, we have discussed dynamical chiral symmetry breaking in
effective chiral quark models of QCD. Because the effective models may contain
loop momentum cut-off as well as non-local interactions, the conserved
axial-vector current is modified accordingly.
Then the low-energy constants and relations, such as the pion decay
constant, the Gell-Mann--Oakes--Renner relation, become very complicated.
Nevertheless, using the CJT action formulation, we have proved 
that the combination of the SD and BS equations preserve the chiral symmetry
and that the BS equation bears the NG pion solution.
It is also shown that the pion decay
constant $f_\pi$ must be defined according to the modified axial-vector
current and that such $f_\pi$ satisfies the PCAC relation.

We have derived a general formula of the pion decay constant in terms of
the quark full propagator and the pion BS amplitude. The formula is consistent
with the loop expansion of the CJT effective action. A numerical analysis given in Sec.\ref{SEC:H100326:3} shows that the consistency of the SD and BS
equations with the chiral symmetry is essential for the low-energy relations.
We have proposed a Pagels--Stokar type formula which gives the pion decay
constant in terms only of the mass function $B(q^2)$ of the SD equation.
We have found that this formula gives a very good approximation and therefore
saves computation time.

 Our intention is to apply the general formulation given in the present paper
to the study of the light $q\bar{q}$ mesons from the chiral symmetry viewpoint.
As we employ realistic effective models, beyond the NJL model, the chiral 
symmetry is not trivially conserved. Thus we need a consistent approach 
of the system of the SD and BS equations. We believe that the present 
formulation gives a consistent view of the dynamical symmetry breaking in the
effective model analyses of low-energy hadrons.

 In a separate paper\cite{TIT01}, we study the pion
in the improved ladder model with finite quark mass. 
The introduction of
 finite quark mass breaks the chiral symmetry explicitly. It is important
 and interesting to investigate how far the chiral symmetry can be applied.
For instance, the Gell-Mann--Oakes--Renner relation is proved from the 
chiral symmetry in the $m_0\to 0$ limit. Its deviation at finite $m_0$
should be studied. Such a study will give an indication to the applicability 
of the chiral perturbation theory.\cite{GL84}\,

The axial $U(1)$ symmetry is known to be broken by the anomaly,
which may be caused by the instanton\cite{Instanton} configration of QCD.
In effective models for quarks, the $U(1)_A$ breaking may be  represented 
by the instanton mediated interaction. In our study of a realistic
model in the flavor $SU(3)$\cite{TIT02},
we employ the Kobayashi--Maskawa--'t Hooft interaction\cite{KM1,TH1},
which consists of a six-quark vertex.
The present general formulation can be easily extended to such six-quark
interaction with an appropriate momentum dependence.
Our formula for the pion decay constant $(\ref{AEQ:H091223:11})$ is
applicable while the proxy of the axial-vector current $(\ref{AEQ:H091223:9})$
requires additional terms from the new interaction.


\section*{Acknowledgement}
 The authors would like to thank Profs. A. G. Williams and K. Kusaka 
for valuable discussions.
 This work is supported in part by the Grant-in-Aid
for scientific research (A)(1) 08304024 and (C)(2) 08640356
of the Ministry of Education, Science and Culture
of Japan.

\end{document}